\documentclass[10pt,amsmath,amssymb,twocolumn,superscriptaddress,nofootinbib,aps,prd,twocolumn,floatfix,showkeys]{revtex4-2}

\usepackage{graphicx}  % needed for figures
\graphicspath{{./plots/}}
\usepackage{bm}        % for math
\usepackage{enumitem}
\usepackage{etoolbox}
\usepackage{tikz-feynman}

\usepackage[colorlinks=true]{hyperref}

\providecommand{\texorpdfstring}[2]{#1}

\newcommand{\comment}[1]{}

\newcommand{\lr}[1]{ \left( #1 \right) }
\newcommand{\lrs}[1]{ \left[ #1 \right] }
\newcommand{\lrc}[1]{ \left\{ #1 \right\} }
\newcommand{\vev}[1]{ \langle \, #1 \, \rangle }

\newcommand{\tr}{ {\rm Tr} \, }

\newcommand{\ket}[1]{ \, | #1 \rangle }
\newcommand{\bra}[1]{ \langle #1 | \, }

\newcommand{\expa}[1]{ \exp{\left( #1 \right)} }
\newcommand{\abs}[1]{| #1 |}

\newcommand{\HH}[0]{\hat{\mathcal{H}}}
\newcommand{\QA}[0]{\hat{\mathcal{Q}}_5}
\newcommand{\JZ}[0]{\hat{\mathcal{J}}_z}

\begin{document}
\sloppy

%It would be also important to mention in Sec. V what values of magnetic flux were used in the simulations (presently only mentioned within the fig. 3 and its caption) and how it was implemented on the lattice with appropriate references. This will make the discussions about the results much more complete.

\title{Out-of-equilibrium Chiral Magnetic Effect from simulations on Euclidean lattices}%\logo}

\author{P. V. Buividovich}
\email{pavel.buividovich@liverpool.ac.uk}
\affiliation{Department of Mathematical Sciences, University of Liverpool, UK}

\date{November 13th, 2024}
\begin{abstract}
The status of the Chiral Magnetic Effect (CME) response in full Quantum Chromodynamics (QCD) has been controversial so far, with previous lattice QCD studies indicating either its strong suppression or vanishing in thermal equilibrium state. We introduce the Euclidean-time correlator of axial charge and electric current as an observable that can be used to study the finite out-of-equilibrium CME response in first-principle lattice QCD simulations with background magnetic field. This observable directly reflects the fact that in the background magnetic field, a state with nonzero axial charge features nonzero electric current. For free fermions, the axial-vector correlator only receives contributions from the Lowest Landau Level, and exhibits a linear dependence on both magnetic field and temperature with a universal coefficient. With an appropriate regularization, non-vanishing axial-vector correlator is compatible with the vanishing of the CME current in thermal equilibrium state with nonzero chiral chemical potential $\mu_5$. We demonstrate that the real-time counterpart of the Euclidean-time axial-vector correlator is intimately related to the real-time form of the axial anomaly equation, which strongly limits possible corrections in full QCD. We present numerical results for the Euclidean-time axial-vector correlator in $SU(2)$ lattice gauge theory with $N_f = 2$ light quark flavours, demonstrating reasonable agreement with free fermion result on both sides of the chiral crossover. The proposed methodology should help to answer the question whether the QCD corrections might be responsible for non-observation of CME in heavy-ion collision experiments such as the RHIC isobar run.
\end{abstract}

\maketitle

\section{Introduction and motivation}
\label{sec:intro}

Since the original proposals which date back almost two decades \cite{Kharzeev:0808.3382,Kharzeev:0711.0950}, the Chiral Magnetic Effect (CME) and, more generally, macroscopic transport responses generated by quantum anomalies became a subject of intense theoretical and experimental studies. Recent non-observation of CME in a dedicated isobar run at RHIC heavy-ion collider \cite{STAR:2109.00131} has generated renewed interest in understanding this phenomenon in Quantum Chromodynamics (QCD). A particularly important question is whether the CME can get significant corrections in QCD, thus leading to its non-observation. In the absence of analytic methods that would be reliable in the low-energy regime of QCD probed in heavy-ion collisions, lattice QCD simulations may be the only way to answer this question from first principles \cite{Buividovich:09:7,Yamamoto:1105.0385}.

\begin{figure}[h!tpb]
\includegraphics[width=0.20\textwidth]{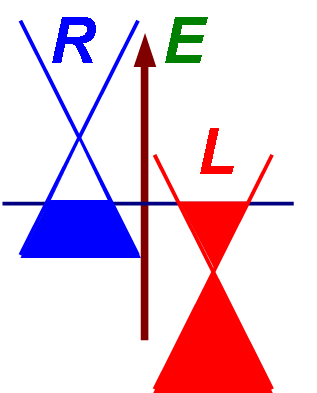}\includegraphics[width=0.20\textwidth]{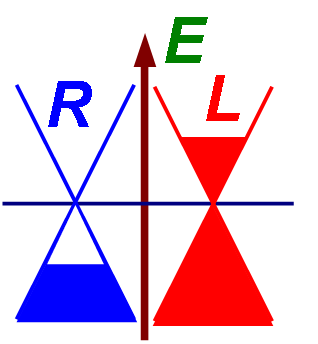}\\
\label{fig:dispersion}
\caption{Schematic dispersion relations and single-particle energy level occupancies for Dirac fermions in thermal equilibrium with nonzero chiral chemical potential $\mu_5$ (\textbf{left}) and with non-equilibrium level occupancy (\textbf{right}).}
\end{figure}

CME is conventionally associated with the renowned formula
\begin{eqnarray}
\label{eq:CME_iconic}
 \vec{j} = \frac{\mu_5}{2 \pi^2} \, \vec{B} ,
\end{eqnarray}
where $\vec{j}$ is the electric current density, $\vec{B}$ is the magnetic field, and $\mu_5$ is the ``chiral chemical potential'' which parameterizes the imbalance of left- and right-handed fermions in the system (for simplicity, here we consider just a single fermion field with unit electric charge). While the electric current and the magnetic field are well-defined physical quantities, most of the controversy around the interpretation of CME is related to the definition of the chiral chemical potential $\mu_5$.

In contrast to $\vec{j}$ and $\vec{B}$, $\mu_5$ is a phenomenological quantity which is not contained in the QCD Lagrangian. It was introduced as an effective description of chirality imbalance created by real-time sphaleron transitions in hot QCD \cite{Kharzeev:0808.3382}.

Quite a few seminal studies of the CME (see e.g. \cite{Kharzeev:0808.3382,Fukushima:1004.2769,Ren:1103.2035} and the lattice studies \cite{Yamamoto:1105.0385,Yamamoto:1111.4681,Braguta:1503.06670}) used the following prescription to introduce the chiral chemical potential into the QCD Hamiltonian $\HH$:
\begin{eqnarray}
\label{eq:Hamiltonian_change0}
 \HH \rightarrow \HH + \mu_5 \QA ,
 \\
 \label{eq:axial_charge_def}
 \QA = \sum\limits_f \int d^3 \vec{x} \, \hat{q}^{\dag}_f\lr{\vec{x}} \gamma_5 \hat{q}_f\lr{\vec{x}}
\end{eqnarray}
where $\QA$ is the axial charge operator and $\hat{q}^{\dag}_f\lr{x}$, $\hat{q}_f\lr{x}$ are the creation/annihilation operators for quarks of flavour $f$. Since the axial charge $\QA$ does not commute with the QCD Hamiltonian $\HH$, the theoretical status of the chiral chemical potential is very different from the one of the baryon chemical potential, or any other chemical potential which corresponds to a conserved charge.

In contrast to chemical potentials which couple to conserved charges, the introduction of $\mu_5$ in (\ref{eq:Hamiltonian_change0}) changes the dispersion relation of Dirac fermions instead of merely shifting the Fermi level. Namely, the energies of left- and right-handed fermions have different dependence on the momentum $\vec{k}$. In a state of thermal equilibrium, the Fermi level remains at zero energy, but the resulting occupation numbers of left- and right-handed fermions become different, as schematically illustrated in Fig.~\ref{fig:dispersion} (left plot). This results in a net chirality imbalance, that is, nonzero expectation value of the axial charge operator $\QA$. 

The prescription (\ref{eq:Hamiltonian_change0}) seemed advantageous at first, as it allowed to consider a state with nonzero axial charge in thermal equilibrium. Furthermore, lattice QCD simulations at finite $\mu_5$ do not suffer from fermionic sign problem \cite{Yamamoto:1105.0385,Yamamoto:1111.4681,Braguta:1503.06670}, in contrast to the case of baryonic chemical potential.

However, it was quite quickly realized that the CME electric current vanishes in a thermal equilibrium state of the Hamiltonian (\ref{eq:Hamiltonian_change0}) with nonzero $\mu_5$ \cite{Rebhan:0909.4782,Buividovich:13:8,Yamamoto:1502.01547,Brandt:2405.09484}, and the finite result (\ref{eq:CME_iconic}), interpreted as thermal expectation value, is a result of an improper regularization of the electric current operator. On the lattice, the use of non-conserved electric current leads to a finite CME signal at finite $\mu_5$\cite{Yamamoto:1105.0385} which is however significantly suppressed in comparison with (\ref{eq:CME_iconic}). 

The CME current can still be nonzero for the Hamiltonian (\ref{eq:Hamiltonian_change0}) in the case of time-dependent magnetic fields \cite{Kharzeev:0907.5007,Yee:0908.4189,Kharzeev:1612.01674,Ren:1103.2035}, which drive chiral fermions out of thermal equilibrium. In principle, this setup can be studied in lattice QCD simulations within the linear response approximation \cite{Landsteiner:1207.5808}. However, even with time-dependent magnetic fields, interpretation of the chiral chemical potential $\mu_5$ in QCD remains unclear.

For the above reasons, first-principle lattice QCD studies of the out-of-equilibrium CME remained inconclusive so far, in contrast to lattice studies of the Chiral Separation and Chiral Vortical effects which both exist in thermal equilibrium \cite{Braguta:1303.6266,Braguta:1401.8095,Buividovich:16:6,Buividovich:20:2,Brandt:2212.02148,Endrodi:2312.02945}.

Within the first-principle QCD description, generation of chirality imbalance (axial charge) is governed by the axial anomaly equation
\begin{eqnarray}
\label{eq:axial_anomaly}
 \frac{d  \vev{\QA}}{d t}
 =
 \frac{\vec{E} \cdot \vec{B}}{2 \pi^2}
 +
 \frac{\vev{\vec{\mathcal{E}}_a \cdot \vec{\mathcal{B}}_a}}{2 \pi^2}
 + \nonumber \\ +
 2 i m_f \int d^3 \vec{x} \, \vev{\hat{q}^{\dag} \gamma_5 \gamma_0 \hat{q}} .
\end{eqnarray}
The net chirality imbalance can be thus created by parallel electric and magnetic fields $\vec{E}$ and $\vec{B}$ acting on QCD matter, or by quantum fluctuations of non-Abelian gauge fields with nonzero scalar product $\vec{\mathcal{E}}_a \cdot \vec{\mathcal{B}}_a$ of chromo-electric and chromo-magnetic fields (sphaleron transitions) \cite{Kharzeev:0808.3382,Berges:1603.03331,Schlichting:1601.07342}. The last term, proportional to the quark mass $m_f$, describes relaxation of axial charge because of explicit breaking of chiral symmetry by the mass term.

\begin{figure}[h!tpb]
 \includegraphics[width=0.40\textwidth]{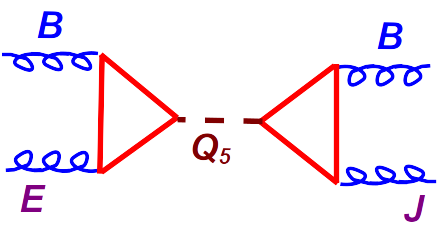}\\
 \label{fig:NMR}
 \caption{A combination of axial anomaly diagrams leading to negative magnetoresistance (quadratic dependence of electric conductivity on magnetic field). In contrast, the lattice observable proposed in this work corresponds to a single fermion diagram with external lines $Q_5$, $\vec{j}$ and $\vec{B}$ (right triangle), see Fig.~\ref{fig:diagrams}.}
\end{figure}

For example, if external electric and magnetic fields are switched on and off sufficiently slowly, they create an excited state with different occupancies for quark states of different chiralities, as illustrated on Fig.~\ref{fig:NMR} (right plot). Note that the dispersion relation of Dirac fermions is not changed, only the occupation numbers are. The resulting state is not a thermal equilibrium state, and can give rise to a nonzero CME current.

Dynamical generation of axial charge can be readily simulated using real-time techniques such as classical statistical field theory \cite{Schlichting:2211.11365,Berges:1603.03331}. Classical-statistical simulations also allow to observe the signatures of the CME on real-time sphaleron field configurations \cite{Mueller:1612.02477,Mueller:1606.00342} and in parallel electric and magnetic fields \cite{Buividovich:16:5}.

However, classical-statistical field theory simulations are based on the classical approximation for gauge field dynamics, and hence cannot provide fully first-principle results. Currently, the only way to obtain first-principle results for out-of-equilibrium transport phenomena in QCD is to use the linear response approximation. The corresponding unperturbed state is the thermal equilibrium state, which can be readily simulated on Euclidean lattices. Results obtained on Euclidean lattices should be then analytically continued back to Minkowski time (or, equivalently, from imaginary Matsubara frequencies to real frequencies) \cite{Meyer:1104.3708}.

A combination of the CME and chirality generation in parallel electric and magnetic fields leads to a specific prediction for the enhancement of electric conductivity in the direction of magnetic field $\vec{B}$, the so-called Negative Magnetoresistance (NMR) \cite{Buividovich:10:1,Landsteiner:1410.6399,Landsteiner:1504.06566,Braguta:1707.09810}. Namely, parallel electric and magnetic fields create a state with nonzero axial charge $\vev{\QA}$. In combination with magnetic field $\vec{B}$, this chirality imbalance creates the CME current as predicted by (\ref{eq:CME_iconic}). This results in a contribution to electric current $\vec{j}$ that is quadratic in magnetic field. If an external electric field is treated within the linear response approximation, this scaling can be readily understood from a combination of two triangular axial anomaly diagrams, as illustrated in Fig.~\ref{fig:NMR}.

NMR is one of the most robust signatures of CME that is especially suitable for lattice QCD simulations within the linear response approximation \cite{Buividovich:10:1,Braguta:1707.09810,Braguta:1910.08516,Braguta:2406.18504} as well as for experimental studies in Dirac semi-metals \cite{Kharzeev:1412.6543}. However, a disadvantage of NMR response is that it is difficult to disentangle the CME contribution from other contributions to DC or AC electric conductivities \cite{Kharzeev:1412.6543}, especially given that quark-gluon plasma is a very good electric conductor \cite{Nikolaev:2008.12326}.

The goal of this paper is to introduce a framework for studying CME in first-principle lattice QCD simulations without the need to add phenomenological terms to QCD Lagrangian, such as the chiral chemical potential $\mu_5$, and without any background contributions that would be nonzero in the absence of magnetic field. Abstracting ourselves from the formula (\ref{eq:CME_iconic}), we can take a broader view on CME and state that CME is a response of electric current density to dynamical fluctuations of net chirality imbalance (axial charge) $\QA$ induced by sphaleron transitions.

On Euclidean lattice, the most natural observable to characterize such a response is the Euclidean-time correlator of the electric current density $\JZ$ with the axial charge $\QA$:
\begin{eqnarray}
\label{eq:corr_general}
 G_{5z}\lr{\tau}
 =
 \vev{
  \QA
  \JZ\lr{\tau}
 } ,
 \quad
 \JZ\lr{\tau}
 =
 e^{-\tau \HH} \JZ e^{\tau \HH} .
\end{eqnarray}
Here the expectation value $\vev{\ldots}$ is the thermal expectation value with respect to the QCD Hamiltonian $\HH$: $\vev{\hat{O}} = \mathcal{Z}^{-1} \tr\lr{\hat{O} e^{-\beta \HH}}$, where $\beta \equiv T^{-1}$ is the inverse temperature $T$ and $\mathcal{Z}$ is the thermal partition function.

We will argue that the correlator (\ref{eq:corr_general}) characterizes the CME strength and is sensitive to non-equilibrium CME response only, in contrast to NMR signal which is measured on the lattice in terms of Euclidean-time correlator of two current operators $\JZ$. At the same time, we will demonstrate that nonzero value of $G_{5z}\lr{\tau}$ does not contradict the vanishing of the CME in thermal equilibrium state with $\mu_5 \neq 0$.

By virtue of Green-Kubo relations \cite{Kadanoff:63:1,Landsteiner:1102.4577}, the Euclidean correlator $G_{5z}\lr{\tau}$ also encodes the real-time CME response, described by the retarded correlator
\begin{eqnarray}
\label{eq:corr_retarded}
 G^R_{5z}\lr{t}
 =
 i \, \theta\lr{t}
 \vev{
  \lrs{\QA\lr{0}, \JZ\lr{t}}
 } ,
 \nonumber \\
 \JZ\lr{t}
 =
 e^{i \HH t} \JZ e^{- i \HH t} ,
\end{eqnarray}
where $\theta\lr{t}$ is the Heaviside step function and $\JZ\lr{t}$ is the time-dependent electric current operator in the Heisenberg representation. Namely, the Fourier transform of the Euclidean-time correlator (\ref{eq:corr_general}) as a function of Matsubara frequencies $w_k = 2 \pi \, T \, k$ is related to analytic continuation of the Fourier transform $G^R_{5z}\lr{w}$ of the real-time retarded correlator (\ref{eq:corr_retarded}):
\begin{eqnarray}
\label{eq:Fourier_defs}
 G_{5z}\lr{w_k} = - G_{5z}^R\lr{w \rightarrow i w_k} ,
 \nonumber \\
 G_{5z}\lr{w_k}
 =
 \int\limits_0^{\beta} d\tau \, e^{i w_k \tau} G_{5z}\lr{\tau} ,
 \nonumber \\
 G_{5z}^R\lr{w} = \int\limits_{-\infty}^{+\infty} dt \, e^{-i w \, t} G^R_{5z}\lr{t} .
\end{eqnarray}

We will show that $G^R_{5z}\lr{t}$ is directly related to the real-time anomaly equation (\ref{eq:axial_anomaly}), which strongly constraints possible corrections to CME. Our main results are equations (\ref{eq:corr_final}) and (\ref{eq:corr_retarded_final}) for $G_{5z}\lr{\tau}$ and $G^R_{5z}\lr{w}$.

In what follows, we will first present a derivation and regularization of the correlator $G_{5z}\lr{\tau}$ for non-interacting, continuum Dirac fermions in the background magnetic field, highlighting some of its interesting features (Sections~\ref{sec:analytics} and \ref{sec:contact}). In Section~\ref{sec:real_time} we will discuss the relation of $G_{5z}\lr{\tau}$ to real-time retarded correlator $G^R_{5z}\lr{t}$, and the relation of $G^R_{5z}\lr{t}$ to the anomaly equation (\ref{eq:axial_anomaly}). We will then present numerical results obtained in $SU\lr{2}$ lattice gauge theory with $N_f = 2$ light dynamical quarks in Section~\ref{sec:lattice}, demonstrating good agreement with the free-fermion result (\ref{eq:corr_final}).

\section{Axial charge - vector current correlator for non-interacting continuum Dirac fermions}
\label{sec:analytics}

In this Section, we will calculate the correlator $G_{5z}\lr{\tau}$ for non-interacting Dirac fermions in the continuum. We start with the non-interacting Dirac Hamiltonian
\begin{eqnarray}
\label{eq:Dirac_Hamiltonian}
 \HH = \int d^3\vec{x} \, \hat{q}^{\dag}\lr{\vec{x}} H \hat{q}\lr{x} ,
 \nonumber \\
 H = \left(
                 \begin{array}{cc}
                  -i \sigma_k \nabla_k   & m \\
                   m                     &   i \sigma_k \nabla_k                 \\
                 \end{array}
 \right)
\end{eqnarray}
in the background of a uniform magnetic field $\vec{B} = \lrc{0, 0, B_z}$, which we assume to be parallel to the $z$ axis. We choose the corresponding Abelian vector gauge field as $A_x = -B_z \, y$, $A_y = A_z = 0$. The system is then translationally invariant in $x$ and $z$ directions. This allows us to use the plain wave basis and replace the partial derivatives $\partial_x$ and $\partial_y$ with the corresponding wave vector components: $\partial_x \rightarrow i \, k_x$ and $\partial_z \rightarrow i \, k_z$. For simplicity, we consider just a single fermionic field $\hat{q}^{\dag}\lr{x}$, $\hat{q}\lr{x}$ with unit electric charge. The terms with the covariant derivatives $\nabla_k$ in the single-particle Dirac Hamiltonian $H$ in (\ref{eq:Dirac_Hamiltonian}) can be represented as
\begin{eqnarray}
\label{eq:Weyl_Hamiltonian}
 -i \sigma_k \nabla_k =
 \left(
                 \begin{array}{cc}
                   k_z                        & k_x + B \, y - \partial_y \\
                   k_x + B \, y + \partial_y  &  -k_z                     \\
                 \end{array}
 \right)
 = \nonumber \\ =
 \left(
                 \begin{array}{cc}
                   k_z                & \sqrt{2 B} \, a^{\dag} \\
                   \sqrt{2 B} \, a  &  -k_z                    \\
                 \end{array}
 \right) .
\end{eqnarray}
Here we introduced the usual magnetic creation and annihilation operators
$a = \frac{1}{\sqrt{2 B}} \lr{k_x + B \, y + \partial_y}$, $a^{\dag} = \frac{1}{\sqrt{2 B}} \lr{k_x + B \, y - \partial_y}$ which satisfy the standard commutation relations $\lrs{a, a^{\dag}} = 1$.

The eigensystem of the single-particle Hamiltonian $H$ consists of:
\begin{itemize}
\item The Lowest Landau Level
\begin{eqnarray}
\label{eq:LLL_Dirac}
\ket{k_z, 0, s} = \left(
   \begin{array}{cc}
       \sqrt{\frac{1}{2} + \frac{k_z}{2 E_s}} \ket{0}   \\
        0 \\
     s \sqrt{\frac{1}{2} - \frac{k_z}{2 E_s}} \ket{0} \\
        0 \\
   \end{array}
 \right)   ,
 \nonumber \\
 E_s\lr{k_z, 0} = s \, \sqrt{m^2 + k_z^2} \, ;
\end{eqnarray}
\item Doubly degenerate higher Landau levels
\begin{eqnarray}
\label{eq:HLL_Dirac}
 \ket{k_z, n, \sigma, s}
 = \nonumber \\ =
 \left(
   \begin{array}{c}
     \frac{1}{2}             \sqrt{\lr{1+\frac{\epsilon_{\sigma}}{E_s}}\lr{1+\frac{k_z}{\epsilon_{\sigma}}}} \, \ket{n} \\
     \frac{\sigma}{2}     \sqrt{\lr{1+\frac{\epsilon_{\sigma}}{E_s}}\lr{1-\frac{k_z}{\epsilon_{\sigma}}}} \, \ket{n-1} \\
     \frac{s}{2}          \sqrt{\lr{1-\frac{\epsilon_{\sigma}}{E_s}}\lr{1+\frac{k_z}{\epsilon_{\sigma}}}} \, \ket{n} \\
     \frac{s\, \sigma}{2} \sqrt{\lr{1-\frac{\epsilon_{\sigma}}{E_s}}\lr{1-\frac{k_z}{\epsilon_{\sigma}}}} \, \ket{n-1} \\
   \end{array}
 \right) ,
 \nonumber \\ {} \nonumber \\
 E_s\lr{k_z, n} = s \sqrt{m^2 + k_z^2 + 2 B \, n} \, ,
\end{eqnarray}
\end{itemize}
where $\sigma = \pm 1$ labels doubly degenerate energy levels $E_s\lr{k_z, n}$, $\epsilon_{\sigma}\lr{k_z} \equiv \sigma \, \sqrt{k_z^2 + 2 B n}$, $s = \pm 1$ labels particle and anti-particle states, and $\ket{n}$, $n = 0, 1, 2, \ldots$ are the harmonic oscillator states generated by the creation/annihilation operators $a^{\dag}$, $a$ from the oscillator ground state $\ket{0}$. For the sake of brevity, we sometimes omit the arguments $k_z$ and $n$ of the eigenenergies. All energy levels are $N_B$ times degenerate, where
\begin{eqnarray}
\label{eq:flux_def}
  N_B = \frac{2 \pi B_z}{L_x \, L_y} \in \mathbb{Z}
\end{eqnarray}
is the integer-valued magnetic flux through the $\lr{x, y}$ plane, and $L_{x,y}$ is the size of the system in $x$ and $y$ directions. With lattice gauge theory simulations in mind, we assume that the Dirac Hamiltonian is defined on a finite torus with sizes $L_x$, $L_y$ and $L_z$ along the three spatial dimensions.

Since the net axial charge $\QA$ as defined in (\ref{eq:axial_charge_def}) is an extensive, volume-averaged quantity, translational invariance allows us to replace the local current density in (\ref{eq:corr_general}) with a volume average
\begin{eqnarray}
\label{eq:jz_q5_single_particle}
\JZ =
\frac{1}{V} \int d^3 \vec{x} \, \hat{q}^{\dag}\lr{\vec{x}} j_z \hat{q}\lr{\vec{x}} ,
\,\,
j_z =
\left(
   \begin{array}{cc}
     \sigma_z & 0 \\
     0   &  -\sigma_z \\
   \end{array}
 \right) ,
\end{eqnarray}
where $V = L_x \, L_y \, L_z$ is the spatial volume. We also introduced the single-particle current operator $j_z$

For non-interacting fermions, the correlator (\ref{eq:corr_general}) can be readily expressed in terms of the single-particle operators $j_z$, $\gamma_5$ as
\begin{eqnarray}
\label{eq:corr_sp}
 G_{5z}^0\lr{\tau}
 =
 \sum\limits_{k,l}
 \bra{k} \gamma_5 \ket{l}
 \bra{l} j_z \ket{k}
 \frac{e^{-\tau E_l}}{1 + e^{-\beta E_l}}
 \frac{e^{-\lr{\beta-\tau} E_k}}{1 + e^{-\beta E_k}} ,
\end{eqnarray}
where the generalized indices $k$ and $l$ label all eigenstates of the single-particle Hamiltonian $H$. We denote these eigenstates as $\ket{k}$, with $E_k$ being the corresponding eigen-energies. In our case, the index $k$ comprises the momentum $k_z$, the Landau level number $n = 0, 1, 2, \ldots$, and the discrete indices $\sigma$ and $s$ in (\ref{eq:LLL_Dirac}) and (\ref{eq:HLL_Dirac}).

In Section~\ref{sec:contact} we will consider a regularization of $G_{5z}\lr{\tau}$. In order to stress that for now we are working with the bare, unregularized observable, in (\ref{eq:corr_sp}) we put a superscript $0$ on $G_{5z}^0\lr{\tau}$.

Taking into account the translational invariance of the system in the direction $z$ and the orthogonality of Landau levels with different values of $n$, we conclude that only states $\ket{k}$ and $\ket{l}$ with the same values of $n$ and $k_z$ can contribute to the correlator (\ref{eq:corr_sp}). To calculate $G_{5z}^0\lr{\tau}$ from (\ref{eq:corr_sp}), we need the matrix elements of $j_z$ and $\gamma_5$ between the single-particle Landau level states (\ref{eq:HLL_Dirac}) and (\ref{eq:LLL_Dirac}). For the higher Landau levels (\ref{eq:HLL_Dirac}) we get:
\begin{eqnarray}
\label{eq:jz_matrix_elements}
 \bra{k_z, n, \sigma_2, s_2} j_z \ket{k_z, n, \sigma_1, s_1}
 =
 \frac{k_z}{E_{s_1}} \delta_{s_1, s_2} \delta_{\sigma_1, \sigma_2}
 + \nonumber \\ +
 \delta_{s_1, -s_2}
 \lr{
   \frac{k_z}{\epsilon_{\sigma_1}} \frac{m}{\abs{E_{s_1}}} \delta_{\sigma_1, \sigma_2}
   +
   \frac{\sqrt{2 B n}}{\abs{\epsilon_{\sigma_1}}} \delta_{\sigma_1, -\sigma_2}
 } ,
\end{eqnarray}
\begin{eqnarray}
\label{eq:q5_matrix_elements}
 \bra{k_z, n, \sigma_1, s_1} \gamma_5 \ket{k_z, n, \sigma_2, s_2}
 = \nonumber \\ =
 \frac{\epsilon_{\sigma_1}}{E_{s_1}} \delta_{s_1, s_2} \delta_{\sigma_1, \sigma_2}
 +
 \delta_{s_1, -s_2} \delta_{\sigma_1, \sigma_2}
 \frac{m}{\abs{E_{s_1}}}
\end{eqnarray}
Using the fact that $s_{1,2}$ and $\sigma_{1,2}$ only take the discrete values $\pm 1$, the product of matrix elements of $\gamma_5$ and $j_z$ which enters the Euclidean correlator (\ref{eq:corr_sp}) can be simplified as
\begin{eqnarray}
\label{eq:q5_jz_product}
 \bra{k_z, n, \sigma_1, s_1} \gamma_5 \ket{k_z, n, \sigma_2, s_2}
 \times \nonumber \\ \times
 \bra{k_z, n, \sigma_2, s_2} j_z \ket{k_z, n, \sigma_1, s_1}
 = \nonumber \\ =
 \sigma_1
 \frac{k_z}{\abs{\epsilon_{\sigma_1}}}
 \delta_{\sigma_1,\sigma_2}
 \lr{
 \delta_{s_1, s_2} \frac{\epsilon_{\sigma_1}^2}{E_{s_1}^2}
 +
 \delta_{s_1, -s_2} \frac{m^2}{E_{s_1}^2}
 } .
\end{eqnarray}
We see that the product of two matrix elements is odd in $\sigma_1$. Since the energies $E_{s_1}\lr{k_z, n}$ and $E_{s_2}\lr{k_z, n}$ and the terms $\abs{\epsilon_{\sigma_1}}$ and $\epsilon_{\sigma_1}^2$ do not depend on $\sigma_1$ or $\sigma_2$, the entire contribution of higher Landau levels under the sum over $k$, $l$ in (\ref{eq:corr_sp}) is odd in $\sigma_1$ and hence vanishes upon summation over $\sigma_1$.

Let us now consider the contribution from the Lowest Landau Level. The matrix elements of $\gamma_5$ and $j_z$ between the Lowest Landau Level states (\ref{eq:LLL_Dirac}) are
\begin{eqnarray}
\label{eq:q5_jz_LLL}
 \bra{k_z, 0, s_1} \gamma_5 \ket{k_z, 0, s_2}
 =
 \delta_{s_1, s_2} \frac{s_1 \, k_z}{\abs{E_{s_1}}}
 +
 \delta_{s_1, -s_2} \frac{m}{\abs{E_{s_1}}} ,
 \nonumber \\
 \bra{k_z, 0, s_2} j_z \ket{k_z, 0, s_1}
 =
 \delta_{s_1, s_2} \frac{s_1 \, k_z}{\abs{E_{s_1}}}
 +
 \delta_{s_1, -s_2} \frac{m}{\abs{E_{s_1}}} .
\end{eqnarray}
The fact that the matrix elements of the single-particle operators of axial charge and vector current $\gamma_5$ and $j_z$ are equal to each other is very instructive and deserves a separate discussion. The expressions (\ref{eq:q5_jz_LLL}) imply that for many-body quantum states with nonzero occupancy of some of the Lowest Landau Levels, electric current is identically equal to the axial charge. Hence any many-body state with nonzero $\vev{\QA}$ which overlaps with the Lowest Landau Level will be also characterized by non-vanishing electric current $\vev{\JZ}$. This is the essence of the CME for non-equilibrium states with different occupancies of left- and right-handed chiral states, as illustrated in Fig.~\ref{fig:dispersion} (right plot).

Let us now proceed with the contribution of the Lowest Landau Level states to $G_{5z}\lr{\tau}$, which is in fact equivalent to two-dimensional axial anomaly by virtue of dimensional reduction in the Lowest Landau Level. The product of the matrix elements (\ref{eq:q5_jz_LLL}) of $\gamma_5$ and $j_z$ that enters (\ref{eq:corr_sp}) can be simplified as
\begin{eqnarray}
\label{eq:q5_jz_LLL_product}
 \bra{k_z, 0, s_1} \gamma_5 \ket{k_z, 0, s_2}
 \bra{k_z, 0, s_2} j_z \ket{k_z, 0, s_1}
 = \nonumber \\ =
 \delta_{s_1, s_2} \frac{k_z^2}{E_{s_1}^2}
 +
 \delta_{s_1, -s_2} \frac{m^2}{E_{s_1}^2} .
\end{eqnarray}
Plugging this expression into (\ref{eq:corr_sp}) and simplifying summations over $s_{1,2}$, we obtain two distinct contributions to $G_{5z}^0\lr{\tau}$:
\begin{eqnarray}
\label{eq:corr_sp_LLL}
 G_{5z}^0\lr{\tau}
 =
 \frac{N_B}{V} \sum\limits_{k_z}
 \sum\limits_{s=\pm1}
 \frac{k_z^2}{E^2} \frac{e^{-\beta \, s \, E}}{\lr{1 + e^{-\beta \, s \, E}}^2}
 + \nonumber \\ +
 \frac{N_B}{V} \sum\limits_{k_z}
 \sum\limits_{s=\pm1}
 \frac{m^2}{E^2} \frac{e^{\lr{\beta/2 - \tau} \, 2 \, s \, E}}{\lr{1 + e^{-\beta \, s \, E}}\lr{1 + e^{+\beta \, s \, E}}} ,
\end{eqnarray}
where we introduced the short-hand notation $E \equiv \abs{E_{s_1}\lr{k_z, 0}} = \sqrt{k_z^2 + m^2}$. The contribution in the first line of (\ref{eq:corr_sp_LLL}) does not depend on $\tau$. The dependence on $\tau$ only comes from the second summand, which is proportional to the square of fermion mass $m$.

Taking the massless limit, we see that only the first line in (\ref{eq:corr_sp_LLL}) gives a nonzero contribution, and $G_{5z}^0\lr{\tau}$ becomes $\tau$-independent. Taking into account that $E = \abs{k_z}$ in this limit, we represent $G_{5z}^0\lr{\tau}$ as
\begin{eqnarray}
\label{eq:corr_sp_LLL_massless_prelim}
 \left. G_{5z}^0\lr{\tau} \right|_{m=0}
 =
 \frac{N_B}{V} \sum\limits_{s=\pm1,k_z}
 \frac{1}{\lr{e^{\beta s E/2} + e^{-\beta s E/2}}^2}
 = \nonumber \\ =
 \frac{N_B}{2 V} \sum\limits_{k_z} \frac{1}{\cosh^2\lr{\beta k_z/2}} .
\end{eqnarray}
For sufficiently large spatial volume, summation over $k_z$ can be replaced with integration as $\sum\limits_{k_z} \rightarrow \frac{L_z}{2 \pi} \int\limits_{-\infty}^{+\infty}d k_z$. Combining this expression with the definition (\ref{eq:flux_def}) of magnetic flux $N_B$, we obtain a very compact and $\tau$-independent result:
\begin{eqnarray}
\label{eq:corr_sp_LLL_massless}
 \left. G_{5z}^0\lr{\tau} \right|_{m=0}
 =
 \frac{B_z}{8 \pi^2} \int\limits_{-\infty}^{+\infty} \frac{d k_z}{\cosh^2\lr{\beta k_z/2}}
 =
 \frac{B_z \, T}{2 \pi^2} .
\end{eqnarray}

Our derivation highlights an important property of the correlator (\ref{eq:corr_general}): it only receives contributions from the Lowest Landau Level states, even at finite fermion mass.

\section{Regularization, contact terms and vanishing CME current at \texorpdfstring{$\mu_5 \neq 0$}{nonzero mu5}}
\label{sec:contact}

An important ingredient in the demonstration that the CME current vanishes in a state of thermal equilibrium with $\mu_5 \neq 0$ is a proper ultraviolet regularization which preserves gauge invariance, for example, the Pauli-Villars regularization. In particular, Pauli-Villars regularization leads to the correct vanishing result for the CME Kubo formulae \cite{Landsteiner:1102.4577,Buividovich:13:8}.

In order to carry out the Pauli-Villars regularization of the correlator $G_{5z}^0\lr{\tau}$, we need to subtract the infinite-mass limit $\left. G_{5z}^0\lr{\tau} \right|_{m \rightarrow +\infty}$ of this correlator from the massless result (\ref{eq:corr_sp_LLL_massless}).

In the limit $m \rightarrow +\infty$, the first, $\tau$-independent term in (\ref{eq:corr_sp_LLL}) is exponentially suppressed and does not contribute. For further analysis, it is convenient to rewrite the second term in (\ref{eq:corr_sp_LLL}) as
\begin{eqnarray}
\label{eq:corr_sp_LLL_tau}
 \left. G_{5z}^0\lr{\tau} \right|_{m \rightarrow +\infty}
 = \nonumber \\ =
 \frac{B_z}{8 \pi^2}
 \int\limits_{-\infty}^{+\infty} d k_z
 \frac{m^2}{m^2 + k_z^2}
 \frac{\cosh\lr{2 E \, \lr{\tau - \beta/2}}}{\cosh^2\lr{\beta E/2}} .
\end{eqnarray}
This representation makes it obvious that $\left. G_{5z}^0\lr{\tau} \right|_{m \rightarrow +\infty}$ is also exponentially suppressed at any $0 < \tau < \beta$, because the $\cosh^2$ term in the denominator will always outgrow the numerator at large energies $E > m$. However, at $\tau = 0$ or $\tau = \beta$ both denominator and numerator grow at the same rate, and exponential suppression is absent. In fact, with $\tau = 0$ and $m \gg \beta$, the ratio $\frac{\cosh\lr{2 E \, \lr{\tau - \beta/2}}}{\cosh^2\lr{\beta E/2}} = \frac{\cosh\lr{\beta \, E}}{\cosh^2\lr{\beta E/2}}$ is exponentially close to $2$ in the entire integration region. Replacing this ratio with $2$, the integral in (\ref{eq:corr_sp_LLL_tau}) becomes very easy to take, and we obtain a linear divergence in the limit $m \rightarrow + \infty$:
\begin{eqnarray}
\label{eq:corr_sp_LLL_tau0_PV}
 \left. G_{5z}^0\lr{\tau = 0} \right|_{m \rightarrow +\infty}
 =
 \frac{B_z}{8 \pi^2} \, 2 \pi \, m + O\lr{e^{-\beta m}} .
\end{eqnarray}
We conclude that $\left. G_{5z}^0\lr{\tau} \right|_{m \rightarrow +\infty}$ is divergent at $\tau = 0$ (identified with $\tau = \beta$), and vanishes at any finite $0 < \tau < \beta$. This suggests that $\left. G_{5z}^0\lr{\tau} \right|_{m \rightarrow +\infty}$ may be proportional to the Dirac $\delta$-function $\delta\lr{\tau}$. In order to check this, let us consider the integral
\begin{eqnarray}
\label{eq:G5z_integral1}
 \int\limits_{0}^{\beta} d\tau \, \left. G_{5z}^0\lr{\tau} \right|_{m \rightarrow +\infty}
 = \nonumber \\ =
 \frac{B_z}{8 \pi^2}
 \int\limits_{-\infty}^{+\infty} d k_z
 \frac{m^2}{m^2 + k_z^2} \, \frac{1}{E} \,
 \frac{\sinh\lr{\beta E}}{\cosh^2\lr{\beta E/2}}
 = \nonumber \\ =
 \frac{B_z}{8 \pi^2}
 \int\limits_{-\infty}^{+\infty} d k_z
 \frac{2 m^2}{\lr{m^2 + k_z^2}^{3/2}}
 =
 \frac{B_z}{2 \pi^2} ,
\end{eqnarray}
where we again used the fact that the ratio $\frac{\sinh\lr{\beta E}}{\cosh^2\lr{\beta E/2}}$ is exponentially close to $2$ in the limit $m \gg \beta$.

We see that the integral of $\left. G_{5z}^0\lr{\tau} \right|_{m \rightarrow +\infty}$ over the entire thermal circle $\tau \in \lrs{0, \beta}$ is finite, as it should be for a delta-function term. We conclude therefore that the Pauli-Villars regularization term can be written as
\begin{eqnarray}
\label{eq:corr_sp_LLL_tau_PV}
 \left. G_{5z}^0\lr{\tau} \right|_{m \rightarrow +\infty}
 =
 \frac{B_z}{2 \pi^2} \, \delta\lr{\tau} .
\end{eqnarray}
Subtracting this regulator term from the massless limit $\left. G_{5z}^0\lr{\tau} \right|_{m=0}$, we obtain our final result for the axial charge-vector current correlator (\ref{eq:corr_general}) for free massless Dirac fermions in the background magnetic field:
\begin{eqnarray}
\label{eq:corr_final}
 \left. G_{5z}\lr{\tau} \right|_{m=0}
 = \nonumber \\ = 
 \left. G_{5z}^0\lr{\tau} \right|_{m=0}
 -
 \left. G_{5z}^0\lr{\tau} \right|_{m \rightarrow + \infty}
 = \nonumber \\ = 
 \frac{B_z \, T}{2 \pi^2} - \frac{B_z \, \delta\lr{\tau}}{2 \pi^2} .
\end{eqnarray}
A remarkable property of this result is that the integral of $\left. G_{5z}\lr{\tau} \right|_{m=0}$ over the thermal circle $\tau \in \lrs{0, \beta}$ vanishes. Let us now demonstrate how this property (and hence the Pauli-Villars regularization) is essential to ensure the vanishing of the CME current in thermal equilibrium state with $\mu_5 \neq 0$.

Let us consider the expectation value $\vev{\JZ} = \mathcal{Z}^{-1} \,\tr\lr{\expa{-\beta \HH - \beta \mu_5 \QA} \, \JZ}$ of electric current in a thermal equilibrium state of the QCD Hamiltonian with nonzero $\mu_5$. It is easy to see that the derivative of $\vev{\JZ}$ with respect to $\mu_5$ at $\mu_5 = 0$ is proportional to $G_{5z}\lr{\tau}$:
\begin{eqnarray}
\label{eq:vanishing_CME2}
 \left. \frac{\partial \vev{\JZ}}{\partial \mu_5} \right|_{\mu_5 = 0}
 = \nonumber \\ =
 \mathcal{Z}^{-1} \,
 \int\limits_0^{\beta}
 d \tau
 \tr\lr{
  e^{-\tau \HH}  \QA e^{-\lr{\beta-\tau} \HH}
  \JZ
  }
  - \nonumber \\ -
 \mathcal{Z}^{-2} \frac{\partial \mathcal{Z}}{\partial \mu_5} \,\tr\lr{e^{-\beta \HH} \, \JZ}
  =
  \int\limits_0^{\beta} d \tau \, G_{5z}\lr{\tau}
  ,
\end{eqnarray}
where we used the fact that $\frac{\partial \mathcal{Z}}{\partial \mu_5} \sim \vev{\QA} = 0$ at $\mu_5 = 0$. Thus the integral $\int\limits_0^{\beta} d \tau \, G_{5z}\lr{\tau}$ should vanish for the CME current to be zero in the thermal equilibrium state with $\mu_5 \neq 0$ (to the linear order in $\mu_5$), as required, for example, by the generalized Bloch theorem \cite{Yamamoto:1502.01547}.

\section{Real-time CME response and relation to axial anomaly}
\label{sec:real_time}

Let us now calculate the retarded correlator (\ref{eq:corr_retarded}) for free non-interacting fermions. A standard expression for the Fourier transform (\ref{eq:Fourier_defs}) within the linear response approximation for non-interacting fermions is
\begin{eqnarray}
\label{eq:G5z_real_time_sp}
 G_{5z}^{R0}\lr{w}
 =
 \sum\limits_{k,l}
 \frac{
  \bra{k} \gamma_5 \ket{l}\bra{l} j_z \ket{k}
 }{
  w - \lr{E_k - E_l} - i \varepsilon
 }
  \times \nonumber \\ \times
 \frac{
   \lr{e^{-\beta E_l} - e^{-\beta E_k}}
 }{
  \lr{1 + e^{-\beta E_k}}
  \lr{1 + e^{-\beta E_l}}
 } ,
\end{eqnarray}
where the superscript $0$ in $G_{5z}^{R0}\lr{w}$ is again used to stress that for now we work with an unregularized quantity. As usual, a small complex shift $w \rightarrow w - i \varepsilon$ ensures the correct pole structure of $G_{5z}^{R0}\lr{w}$ in the plane of complex $w$.

Similarly to the case of Euclidean correlator $G_{5z}\lr{\tau}$, it is clear that higher Landau levels with the product of matrix elements $\bra{k} \gamma_5 \ket{l}\bra{l} j_z \ket{k}$ given by (\ref{eq:q5_jz_product}) will not contribute to the real-time propagator (\ref{eq:G5z_real_time_sp}), as contributions of states with different $\sigma$ labels cancel each other.

Plugging in the expression (\ref{eq:q5_jz_LLL_product}) for the product of matrix elements $\bra{k} \gamma_5 \ket{l}\bra{l} j_z \ket{k}$ between the Lowest Landau Level states, we again conclude that only the states $\ket{k} = \ket{k_z, 0, s_1}$ and $\ket{l} = \ket{k_z, 0, -s_1}$ contribute to $G^{R0}_{5z}\lr{w}$ at $w \neq 0$, yielding the contribution
\begin{eqnarray}
\label{eq:G5z_real_time_LLL1}
 G_{5z}^{R0}\lr{w}
 =
 \sum\limits_{s=\pm1, k_z}
 \frac{
  m^2
 }{
  k_z^2 + m^2
 }
 \frac{
  e^{\beta \, s \, E} - e^{-\beta \, s \, E}
 }{w - 2 s E - i \varepsilon}
 \times \nonumber \\ \times
 \frac{
   1
 }{
  \lr{1 + e^{-\beta s E}}
  \lr{1 + e^{ \beta s E}}
 } ,
\end{eqnarray}
where again $E = +\sqrt{k_z^2 + m^2}$. After some manipulations with exponentials $e^{\pm \beta \, s \, E}$ in the numerator and denominator of (\ref{eq:G5z_real_time_LLL1}), the expression (\ref{eq:G5z_real_time_LLL1}) can be rewritten as
\begin{eqnarray}
\label{eq:G5z_real_time_LLL2}
 G_{5z}^{R0}\lr{w}
 =
 \sum\limits_{s=\pm1, k_z}
 \frac{
  m^2
 }{
  k_z^2 + m^2
 }
 \frac{
  \tanh\lr{\beta \, s \, E/2}
 }{w - 2 s E - i \varepsilon}
 = \nonumber \\ =
 \sum\limits_{k_z}
 \frac{
  4 m^2
 }{
  \sqrt{k_z^2 + m^2}
 }
 \frac{
  \tanh\lr{\beta \sqrt{k_z^2 + m^2}/2}
 }{\lr{w - i \varepsilon}^2 - 4 m^2 - 4 k_z^2}
 .
\end{eqnarray}
Counting the powers of $k_z$ in the denominator, it becomes clear that summation over $k_z$ is UV finite for any finite $m$ and $w$. At any finite $m$, $G_{5z}^{R0}\lr{w}$ has singularity at $w = 2 m$. A direct calculation also shows that the unregularized propagator $G_{5z}^{R0}\lr{w}$ vanishes in the chiral limit $m \rightarrow 0$.

However, similarly to the case of Euclidean correlator $G_{5z}^0\lr{\tau}$, the retarded correlator (\ref{eq:G5z_real_time_LLL2}) still needs to be regularized by subtracting the contribution of the Pauli-Villars terms. To this end, we consider $G_{5z}^{R0}\lr{w}$ in the infinite mass limit $m \rightarrow +\infty$. At $m \gg \beta$, the function $\tanh\lr{\beta \, \sqrt{k_z^2 + m^2}/2}$ in (\ref{eq:G5z_real_time_LLL2}) is exponentially close to one, and we can simplify (\ref{eq:G5z_real_time_LLL2}) as
\begin{eqnarray}
\label{eq:G5z_real_time_PV1}
 \left.G_{5z}^{R0}\lr{w}\right|_{m \rightarrow +\infty}
 = \nonumber \\ =
 \sum\limits_{k_z}
 \frac{
  4 m^2
 }{
  \sqrt{k_z^2 + m^2}
 }
 \frac{1}{\lr{w - i \varepsilon}^2 - 4 m^2 - 4 k_z^2}
 .
\end{eqnarray}
We now replace $\sum\limits_{k_z} \rightarrow \frac{L_z}{2 \pi} \int\limits_{-\infty}^{+\infty} d k_z$, and remember that the Lowest Landau Level is $N_B$ times degenerate, with $N_B = \frac{B_z \, L_x \, L_y}{2 \pi}$. We then obtain
\begin{eqnarray}
\label{eq:G5z_real_time_PV2}
 \left.G_{5z}^{R0}\lr{w}\right|_{m \rightarrow +\infty}
 = \nonumber \\ =
 \frac{B_z}{4 \pi^2}
 \int\limits_{-\infty}^{+\infty}
 \frac{d k_z}{\lr{w - i \varepsilon}^2 - 4 m^2 - 4 k_z^2}
 \frac{
  4 m^2
 }{
  \sqrt{k_z^2 + m^2}
 }
 .
\end{eqnarray}
The integral can now be taken analytically at any $m$. In the limit $m \rightarrow +\infty$, we get a very simple result, similar in spirit to (\ref{eq:corr_sp_LLL_tau_PV}):
\begin{eqnarray}
\label{eq:G5z_real_time_PV3}
 \left.G_{5z}^{R0}\lr{w}\right|_{m \rightarrow +\infty}
 = - \frac{B_z}{2 \pi^2}.
\end{eqnarray}
Remembering that the unregularized massless limit $\left.G_{5z}^{R0}\lr{w}\right|_{m \rightarrow +\infty}$ was equal to zero, we conclude that the massless limit of the regularized retarded correlator is
\begin{eqnarray}
\label{eq:corr_retarded_final}
 \left.G_{5z}^{R}\lr{w}\right|_{m \rightarrow 0}
 = \nonumber \\ =
 \left.G_{5z}^{R0}\lr{w}\right|_{m \rightarrow 0}
 -
 \left.G_{5z}^{R0}\lr{w}\right|_{m \rightarrow +\infty}
 = \frac{B_z}{2 \pi^2}.
\end{eqnarray}

It is remarkable that in the massless limit, the only nonzero contribution to the retarded correlator $G_{5z}^R$ comes from the Pauli-Villars regulator term. Our derivation as well as the analytic relations (\ref{eq:Fourier_defs}) between the Euclidean and real-time correlators make it clear that the nonzero result (\ref{eq:corr_retarded_final}) is in one-to-one correspondence with the contact term proportional to $\delta\lr{\tau}$ in (\ref{eq:corr_final}).

Let us now clarify the relation of the result (\ref{eq:corr_retarded_final}) to the conventional CME result (\ref{eq:CME_iconic}) as well as to the axial anomaly equation (\ref{eq:axial_anomaly}). This relation can already be seen at the level of the diagrammatic representation in the top of Fig.~\ref{fig:diagrams}.

Within the linear response approximation, the retarded correlator (\ref{eq:corr_retarded}) describes the response of the time-dependent expectation value $\vev{\JZ\lr{t}}$ of electric current to a small time-dependent perturbation of the Hamiltonian $\HH$ by the axial charge operator $\QA$: $\HH\lr{t} \rightarrow \HH + \mu_5\lr{t} \QA\lr{t}$:
\begin{eqnarray}
\label{eq:retarded_response}
 \vev{\JZ\lr{t}} = \int\limits_{-\infty}^t dt' G^R_{5z}\lr{t - t'} \mu_5\lr{t'} = \frac{\mu_5\lr{t} \, B_z}{2 \pi^2} ,
\end{eqnarray}
where we used the fact that the inverse Fourier transform of $G^R_{5z}\lr{w}$ given by (\ref{eq:corr_retarded_final}) is simply $G^R_{5z}\lr{t} = \frac{B_z \, \delta\lr{t}}{2 \pi^2}$. While this result is simply the standard CME formula (\ref{eq:CME_iconic}), our derivation clarifies the role of $\mu_5\lr{t}$ as that of a \emph{time-dependent perturbation}, in contrast to the assumption of the thermal equilibrium state with time-independent $\mu_5$.

As a side remark, let us note that the retarded correlator (\ref{eq:corr_retarded_final}) is real-valued and has no pole singularities at finite frequencies $w$. This implies that the spectral function $S_{5z}\lr{w} = \frac{1}{\pi} {\mathrm Im} G^R_{5z}\lr{w}$ vanishes for all $w$, and time-dependent $\mu_5\lr{t}$ does not perform work. In contrast, for the correlator of two vector currents, the spectral function is finite and proportional to AC conductivity.

To discuss the relation to the real-time anomaly equation (\ref{eq:axial_anomaly}), let us consider the complex-conjugate retarded correlator $\bar{G}^R_{z5}\lr{t}$, which can be rewritten as
\begin{eqnarray}
\label{eq:Gz5_retarded}
 \bar{G}^R_{z5}\lr{t} = i \, \theta\lr{t}
 \vev{
  \lrs{\JZ\lr{0}, \QA\lr{t}}
 } ,
 \nonumber \\
 \QA\lr{t}
 =
 e^{i \HH t} \QA e^{- i \HH t} .
\end{eqnarray}
Comparing this expression with (\ref{eq:retarded_response}), we can interpret $\bar{G}^R_{z5}\lr{t}$ as a response of time-dependent axial charge expectation value $\vev{\QA\lr{t}}$ to a small perturbation of the Hamiltonian by the electric current operator $\JZ\lr{t}$, $\HH\lr{t} \rightarrow \HH + A_z\lr{t} \JZ\lr{t}$. However, the coefficient $A_z\lr{t}$ in the perturbed Hamiltonian $\HH\lr{t}$ is nothing but the $z$ component of the electromagnetic vector gauge field. Similarly to (\ref{eq:retarded_response}), we can write
\begin{eqnarray}
\label{eq:retarded_response_conj}
 \vev{\QA\lr{t}} = \int\limits_{-\infty}^t dt' \bar{G}^R_{5z}\lr{t - t'} = \frac{A_z\lr{t} \, B_z}{2 \pi^2} ,
\end{eqnarray}
Differentiating this expression over time $t$ and remembering that $\frac{d A_z\lr{t}}{d t} = E_z\lr{t}$, where $E_z\lr{t}$ is the electric field parallel to the magnetic field, we arrive precisely at the anomaly equation (\ref{eq:axial_anomaly}) in the massless limit $m \rightarrow 0$. The term responsible for sphaleron transitions is of course absent as we work with non-interacting fermions.

\begin{figure}[h!tpb]
\includegraphics[width=0.48\textwidth]{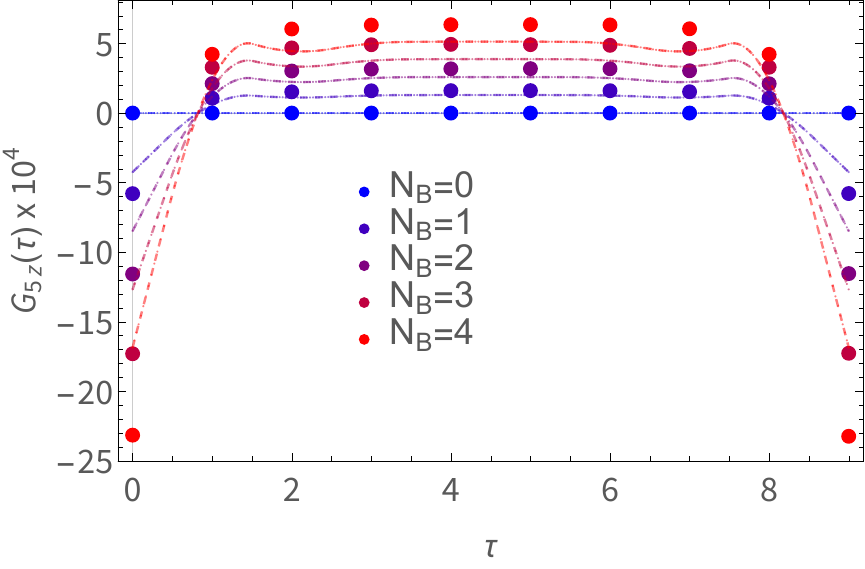}\\
\label{fig:QAJV_vs_t}
\caption{Axial charge-vector current correlator $G_{5z}\lr{\tau}$ (\ref{eq:corr_general}) in $SU\lr{2}$ lattice gauge theory with $N_f=2$ light dynamical quarks on $30^3 \times 10$ lattice with magnetic fluxes $N_B = 0 \ldots 4$. Dashed lines correspond to free fermion results obtained on the same lattice. All values are given in lattice units.}
\end{figure}

\begin{figure}
\begin{tikzpicture}
\begin{feynman}
\vertex[draw,circle] (Q) at (0,0) {$\mathcal{Q}_5$};
\vertex[draw,circle] (Jz) at (5,0) {$\mathcal{J}_z$};
\vertex[draw,circle,blue] (B1) at (2.5,3) {$B$};
\vertex (B0) at (2.5,4);
\vertex (g1) at ($(Q)!0.33!(B1)$);
\vertex (g2) at ($(Q)!0.33!(Jz)$);
\vertex (g3) at ($(B1)!0.33!(Jz)$);
\vertex (g4) at ($(B1)!0.66!(Jz)$);
\vertex (g5) at ($(Q)!0.66!(B1)$);
\vertex (g6) at ($(Q)!0.66!(Jz)$);
\vertex (g7) at ($(g5)!0.33!(g6)$);
\vertex (g8) at ($(g5)!0.66!(g6)$);
\diagram* {
        (Q) -- (g1) -- (B1) -- (Jz) -- (Q),
        (B1) -- [dashed,blue] (B0),
        (g1) -- [gluon] (g2),
        (g3) -- [gluon,half left] (g4),
        (g5) -- [gluon] (g7) -- [half left,magenta] (g8) -- [gluon] (g6),
        (g8) -- [half left,magenta] (g7)
      };
\end{feynman}
\end{tikzpicture}\\
\begin{tikzpicture}
\begin{feynman}
\vertex[draw,circle] (Q) at (0,0) {$\mathcal{Q}_5$};
\vertex[draw,circle] (Jz) at (3,0) {$\mathcal{J}_z$};
\vertex (g1) at (0,2);
\vertex (g2) at (3,2);
\vertex (g3) at (0,1);
\vertex (g4) at (3,1);
\vertex (g5) at (0,3);
\vertex (g6) at (3,3);
\vertex[draw,circle,blue] (B1) at (1.5,4.5) {$B$};
\vertex (B0) at (1.5,5.5);
\diagram* {
        (Q) -- (g3) -- (g1) -- (g2) -- (g4) -- (Jz) -- (Q),
        (g3) -- [gluon] (g4),
        (g1) -- [gluon] (g5),
        (g2) -- [gluon] (g6),
        (g5) -- [magenta](g6) -- [magenta](B1) -- [magenta](g5),
        (B1) --[dashed,blue] (B0)
      };
\end{feynman}
\end{tikzpicture}\\
\caption{Fermionic diagrams that contribute to the CME observable $G_{5z}\lr{\tau}$. \textbf{Above:} the leading connected diagram contributing to the linear dependence on the magnetic field $B$. \textbf{Below:} one of the sub-leading contributions from disconnected fermionic diagrams that is linear in $B$.\\
Any virtual gluon lines within the outer fermionic loops connecting $\mathcal{Q}_5$, $\mathcal{J}_z$ and $B$ vertices can be omitted, or more gluon lines (also with virtual fermion loop insertions) can be added. They are just shown to illustrate how gluon lines will dress the propagators of valence quarks. Valence quarks correspond to outer fermion loops.}
\label{fig:diagrams}
\end{figure}

\section{Lattice results}
\label{sec:lattice}

An important question in the context of experimental detection of CME in heavy-ion collision experiments \cite{STAR:2109.00131} is how strongly does the Euclidean axial-vector correlator $G_{5z}\lr{\tau}$ in full QCD deviate from the free fermion result (\ref{eq:corr_final}).

As a preliminary step towards full lattice QCD simulations, in this work we calculate $G_{5z}\lr{\tau}$ in $SU\lr{2}$ lattice gauge theory with $N_f = 2$ light dynamical quarks. We use the same ensembles of gauge field configurations generated with dynamical staggered fermions on $30^3 \times N_{\tau}$ lattices as in \cite{Buividovich:20:1,Buividovich:20:2,Buividovich:21:1}. Fermionic correlators are calculated used Wilson-Dirac valence quarks with background magnetic field. Our previous studies of axial-vector current-current correlators \cite{Buividovich:20:2} on the same ensemble of gauge configurations suggests that Wilson-Dirac valence quarks produce results that are in good agreement with Domain Wall fermion calculations. At the same time, mixed-action lattice simulations with valence Domain Wall fermions and staggered sea quarks are known to have reasonably small mixed-action artifacts of the order of few percents \cite{Berkowitz:1701.07559}.

We consider a subset of configurations with zero chemical potential and temporal lattice sizes $N_{\tau} = 4, 6, 8, 10, 12, 14, 16, 18, 20, 22$. With fixed gauge coupling $\beta_g = 1.7$ (and hence fixed lattice spacing $a$), the chiral/deconfinement crossover happens around $N_{\tau} = 16$. 

In this proof-of-principle study, magnetic field is not included in the action of staggered sea quarks. We also neglect the contribution of disconnected fermionic diagrams in $G_{5z}\lr{\tau}$. To the linear order in magnetic field strength $B$, both contributions are described by diagrams similar to the second diagram in Fig.~\ref{fig:diagrams}. Similar box-type diagrams appear in hadronic light-by-light scattering amplitudes and are known to be strongly suppressed in comparison with the leading contribution from connected fermionic diagrams (first diagram in Fig.~\ref{fig:diagrams}) \cite{Aoyama:2006.04822}.

\begin{figure}[h!tpb]
\includegraphics[width=0.48\textwidth]{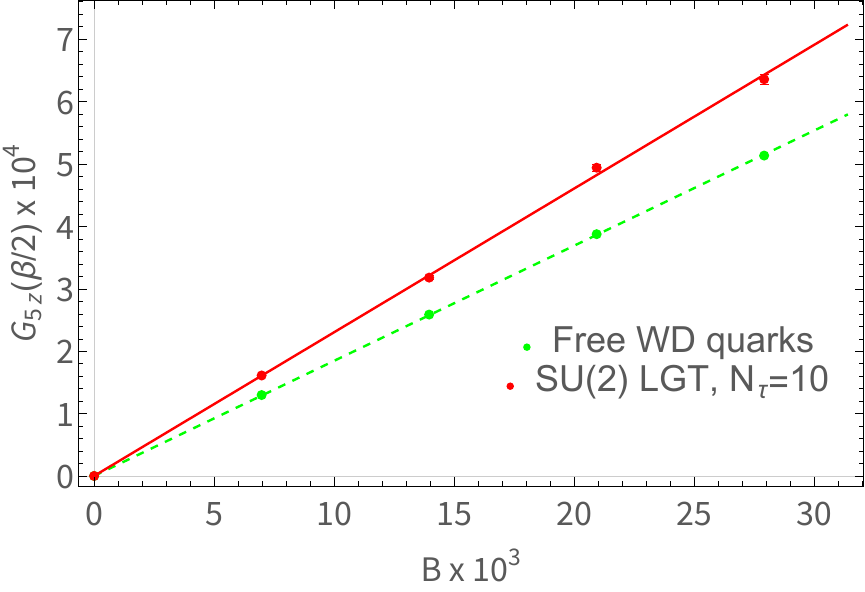}\\
\label{fig:Slope}
\caption{Plateau height $G_{5z}\lr{\beta/2}$ as a function of magnetic field strength, along with a linear fit (solid line).}
\end{figure}

For the electric current operator $\JZ$, we use the conserved lattice vector current for Wilson-Dirac fermions \cite{Buividovich:20:1,Buividovich:21:1}. For the axial charge $\QA$, we use the point-split definition of the axial current, proposed for Wilson-Dirac fermions in \cite{Rakow:1511.05304,Rakow:1612.04992}. The axial charge is not conserved for Wilson-Dirac fermions, and hence gets renormalized. 

The corresponding multiplicative renormalization constant $Z_A^{WD}$ can be estimated based on the results presented in \cite{Buividovich:20:2}, where correlators of vector and axial currents were calculated with the exactly the same lattice action as in the present paper, and with both Wilson-Dirac and Domain Wall fermions. The kinematics was different, but this difference is irrelevant for multiplicative renormalization of the axial current operator. Since Domain Wall fermions are nearly chiral, one can assume that the corresponding axial charge renormalization constant $Z_A^{DW}$ is much closer to unity than for Wilson-Dirac fermions. We can then estimate $Z_A^{WD}$ from the ratio of axial-vector correlators calculated using Wilson-Dirac and Domain Wall fermions. Taking the ratios of the data points on Fig.~3 in \cite{Buividovich:20:2} (top right plot, where statistical errors are smallest) and performing a linear extrapolation to zero lattice momentum, we obtain $Z_A^{WD} = 1.10 \pm 0.05$. Statistical error in this result is dominated by the results for Domain Wall fermions, and are much larger than the statistical error in our numerical data obtained with Wilson-Dirac fermions. To avoid cluttering the plot with additional error bars which are not statistically independent, we multiply all our raw lattice data for $G_{5z}\lr{\tau}$ by $Z_A^{WD} = 1.10$, but do not modify the error bars (as any statistical error in $Z_A$ will lead to a 100\% correlated additional error in all our results). As expected from perturbative analysis \cite{Rakow:1612.04992,Constantinou:1610.06744}, deviations of $Z_A^{WD}$ from unity are small and comparable with our finite-volume and finite-spacing artifacts. It is known that these deviations vanish in the continuum limit \cite{Constantinou:1610.06744}. 

\begin{figure}[h!tpb]
\includegraphics[width=0.47\textwidth]{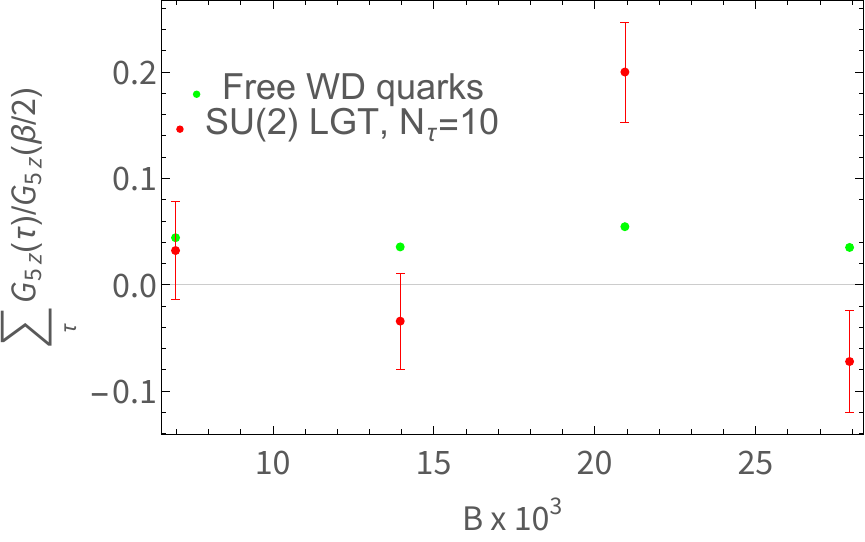}\\
\label{fig:CMET}
\caption{$\sum_{\tau} G_{5z}\lr{\tau}$ on $30^3 \times 10$ lattice as a function of magnetic field strength $B$, normalized to the plateau height $G_{5z}\lr{\beta/2}$.}
\end{figure}

Lattice results for the correlator $G_{5z}\lr{\tau}$ are shown on Fig.~\ref{fig:QAJV_vs_t} for different values of the total magnetic flux $N_B = 0 \ldots 4$. For comparison, free fermion result on the same lattice is also shown. While we only present such a plot for $N_{\tau} = 10$, the data for other temporal lattice sizes looks very similar. We observe a good agreement with the free-fermion result (\ref{eq:corr_final}): large negative values at $a \, \tau = 0$ and $a \, \tau = N_{\tau}-1$, and a characteristic plateau at intermediate values of $\tau$. We define the height of the plateau as $G_{5z}\lr{\beta/2}$, and find that it is proportional to magnetic flux $N_B$, or, equivalently, to magnetic field strength. This proportionality is illustrated on Fig.~\ref{fig:Slope}. On Fig.~\ref{fig:CMET} we also demonstrate that the sum of all correlator values $\sum\limits_{\tau} G_{5z}\lr{\tau}$ is close to zero within statistical errors, and is also significantly smaller than the plateau height $G_{5z}\lr{\beta/2}$. This is in perfect agreement with the vanishing of $\int\limits_0^{\beta} d\tau \, G_{5z}\lr{\tau}$ according to our final result (\ref{eq:corr_final}).

Interestingly, Pauli-Villars regularization seems completely unnecessary for the equation (\ref{eq:corr_sp_LLL}), where all integrals are UV-finite. However, we see that these regulator terms are actually necessary in order to match the lattice data. An explanation of how lattice fermions reproduce the Pauli-Villars regulator terms will be presented elsewhere.

\begin{figure}[h!tpb]
\includegraphics[width=0.47\textwidth]{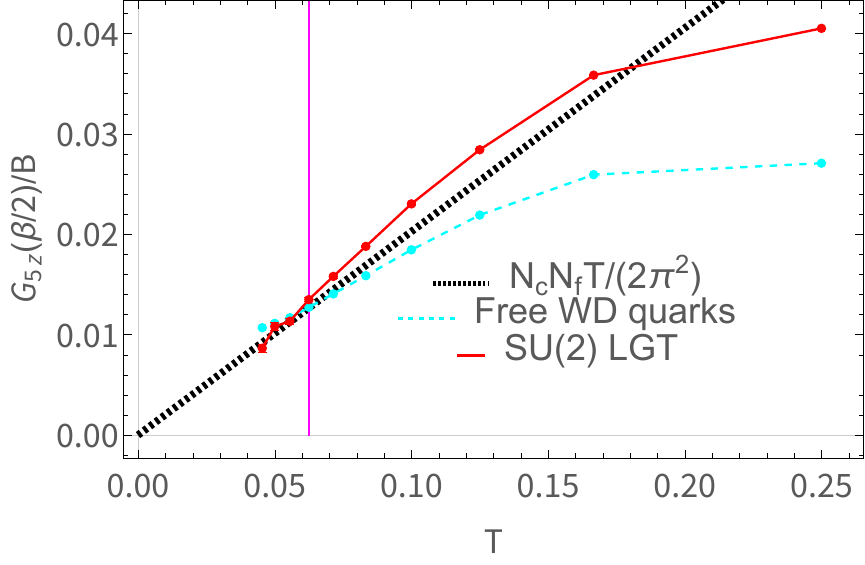}\\
\label{fig:QAJV_vs_T}
\caption{The slope of the linear dependence of $G_{5z}\lr{\beta/2}$ on the magnetic field $B$ as a function of temperature (in lattice units), compared with the free-fermion continuum result $G_{5z}\lr{\beta/2}/B = N_c \, N_f \, T/\lr{2 \pi^2}$ and the corresponding result for free Wilson-Dirac fermions. Vertical magenta line indicates the position of the chiral crossover at $T_c^{-1} = N_{\tau} = 16$ (all quantities in lattice units).}
\end{figure}

We further extract the slope of the linear dependence of the plateau value $G_{5z}\lr{\beta/2}$ on magnetic field $B$ using a linear fit, as illustrated on Fig.~\ref{fig:Slope}. Denoting the slope as $G_{5z}\lr{\beta/2}/B$, we plot it on Fig.~\ref{fig:QAJV_vs_T} as a function of temperature $T$ (in lattice units). For comparison, we also plot the slope $G_{5z}\lr{\beta/2}/B = \frac{N_c \, N_f \, T}{2 \pi^2}$ for our free fermion result (with the factor of $N_c \, N_f$ taking into account the contributions of $N_c = 2$ color states and $N_f = 2$ flavours).

Even though the pion mass is rather large for our lattice ensemble ($m_{\pi}/m_{\rho} = 0.4$), the agreement with the free fermion result (\ref{eq:corr_final}) is reasonably good. Deviation from the linear scaling $G_{5z}\lr{\beta/2}/B = \frac{N_c \, N_f \, T}{2 \pi^2}$ for $N_{\tau} = 4$ is likely to be the lattice artifact, although this needs to be checked by a proper continuum extrapolation. Remarkably, the slope $G_{5z}\lr{\beta/2}/B$ appears to be completely insensitive to the chiral crossover at $N_{\tau} = 16$.

\section{Discussion and conclusions}
\label{sec:discussion}

We introduced the Euclidean-time axial-vector correlator $G_{5z}\lr{\tau}$ (\ref{eq:corr_general}) as a lattice observable that is able to capture the out-of-equilibrium Chiral Magnetic Effect (CME) in first-principle lattice QCD simulations with background magnetic field. This observable should allow to characterize corrections to CME in full QCD in a meaningful and quantitative way, potentially allowing for comparison with experimental data from heavy-ion colliders (e.g. the RHIC isobar run). In the context of anomalous transport, so far QCD corrections were only systematically considered for the Chiral Separation Effect (CSE) \cite{Buividovich:16:6,Buividovich:20:2,Brandt:2212.02148,Endrodi:2312.02945}.

Our preliminary numerical results presented in Section~\ref{sec:lattice} suggest that deviations of the CME response from the free fermion result (\ref{eq:corr_final}) are not very large in finite-temperature gauge theories with dynamical fermions. This finding is in sharp contrast with previous lattice studies, where the use of lattice observables different from $G_{5z}\lr{\tau}$ led to conclusions that the CME is either strongly suppressed in comparison with the free fermion result  \cite{Yamamoto:1105.0385} or vanishes altogether \cite{Brandt:2405.09484}. 

We use linear response approximation to treat the CME current as a small response to small non-equilibrium fluctuations of axial charge on top of the QCD thermal equilibrium state. This is a standard approximation used to extract non-equilibrium transport coefficients (electric conductivity, viscosity) from first-principle lattice QCD simulations. Note that this linear response setup is different from the studies of CME in far-from-equilibrium, quickly evolving state of QCD matter created immediately after heavy-ion collisions \cite{Mueller:1612.02477,Mueller:1606.00342}.  

All quantities entering $G_{5z}\lr{\tau}$ are well-defined and well-understood on the lattice \footnote{The axial charge is not uniquely defined for non-chiral lattice fermions (such as the Wilson-Dirac fermions used in this work), however, the renormalization of axial current for Wilson fermions is very well understood \cite{Rakow:1511.05304,Rakow:1612.04992}.}. The expectation value is calculated with respect to the conventional thermal equilibrium state without any artificial extra terms like the chiral chemical potential. Fluctuations of axial charge (net chirality) that generate the CME current are thermal in nature, as indicated by the linear dependence of $G_{5z}\lr{\tau}$ on temperature.

The intention of this paper is to suggest a new lattice observable rather than present a high-precision lattice calculation of CME. There are a few further technical improvements that are necessary for the full lattice QCD calculation:
\begin{itemize}
 \item Inclusion of background magnetic field into the sea quark action.
 \item Inclusion of disconnected fermionic diagrams.
 \item Proper continuum and large volume extrapolations.
 \item Proper renormalization of the axial charge operator (likely to be at the level of few percents).
\end{itemize}
Our free-fermion result (\ref{eq:corr_final}) is a reference result to compare with numerical lattice QCD data.

The real-time counterpart of $G_{5z}\lr{\tau}$ is the retarded correlator $G^R_{5z}\lr{t}$ defined in (\ref{eq:corr_retarded}). Green-Kubo relations (\ref{eq:Fourier_defs}) express $G_{5z}$ in Matsubara frequency space in terms of the analytic continuation of $G^R_{5z}$ from real to imaginary frequencies.

The relation of $G^R_{5z}\lr{t}$ to the axial anomaly equation (\ref{eq:axial_anomaly}), discussed in Section~\ref{sec:real_time}, suggests that in full QCD, the Euclidean correlator $G_{5z}\lr{\tau}$ will only receive corrections from the last two terms of the anomaly equation (\ref{eq:axial_anomaly}), that is, from sphaleron transitions and from finite-mass corrections. While finite-mass corrections are already included in our lattice results and appear to be reasonably small, a quantitative study of the contribution of $\vec{\mathcal{E}}_a \cdot \vec{\mathcal{B}}_a$ term would require the inclusion of magnetic field into the sea quark action as well as analysis of disconnected fermionic diagrams.

\begin{acknowledgements}
This work has been supported by STRONG-2020 "The strong interaction at the frontier of knowledge: fundamental research and applications” which received funding from the European Union’s Horizon 2020 research and innovation programme under grant agreement No 824093.
This work was funded in part by the STFC Consolidated Grant ST/T000988/1. Numerical simulations were undertaken on Barkla, part of the High Performance Computing facilities at the University of Liverpool, UK.
$SU\lr{2}$ gauge field configurations used in this work were generated in collaboration with Dominik Smith and Lorenz von Smekal using the GPU cluster at the Institute for Theoretical Physics at Giessen University.
The author is grateful to Andrey Kotov and Bastian Brandt for stimulating discussions which motivated this work.
\end{acknowledgements}

%\bibliography{Buividovich}
%\bibliographystyle{apsrev}

\end{document}